\documentclass{cimento}
    \title{Single top: prospects at LHC}
    \author{ M.~Cristinziani\from{ins:bonn}, 
             G.~Petrucciani\from{ins:pisa}}
    \instlist{\inst{ins:bonn}Physikalisches Institut, Universit\"at
    Bonn, Germany
              \inst{ins:pisa}Scuola Normale Superiore di Pisa and INFN Sezione di Pisa, Pisa, IT}

    \PACSes{\PACSit{14.65.Ha}{Top quarks}}
\begin{document}

\maketitle

\begin{abstract}
Single top quark processes are interesting as direct probes of the $Wtb$ vertex, and are also an important background in searches of the Higgs
boson and beyond the standard model physics.
Both ATLAS and CMS have performed studies with simulated data to estimate the expected uncertainty on the production cross section measurements 
of the three single top processes ($t$-channel, $s$-channel, $tW$ production) in the first years of LHC operations. Results in the
different channels and for the two experiments are reported.
\end{abstract}

\section{Introduction}
At LHC the dominant mode for top quark production is in $t\bar{t}$ pairs through strong interactions, 
gluon fusion and  $q\bar{q}$ annihilation. The cross section predicted from NLL calculations is 
$833^{+52}_{-39}$ pb\cite{xsec-tt}.

Production of single top quarks through the electroweak $Wtb$ vertex is also possible, in three different modes.
The most important process is $t$-channel production, where a time-like W boson is exchanged between a light quark 
and a bottom quark from a nearly collinear splitting $g \to b\bar{b}$; the expected cross section at NLO for LHC is 
240 pb\cite{xsec-tch}. For the first measurements only events with $W\to\ell\nu$ ($\ell = e,\mu$) will be usable, 
which reduces the effective cross section to 26 pb for each lepton flavor.
 
The associated production of a top quark and a $W$ boson has a sizeable expected cross section at LHC, 66 pb\cite{xsec-wt}, 
while it is expected to be invisible at Tevatron ($\sigma \sim 0.2$ pb); the decay modes of greater experimental 
interest are the semi-leptonic (one $W$ decaying to $\ell\nu$, the other into a pair of hadronic jets), 
with a $\sigma \cdot BR \simeq 0.4$ pb for each lepton flavor; and the di-leptonic with opposite flavor leptons 
($tW\to b e\mu\nu_e\nu_\mu$).

The $s$-channel top production has the smallest expected cross section at LHC, 10 pb\cite{xsec-sch}; 
it proceeds via $q\bar{q}'$ annihilation in an off-shell 
$W$ boson with a $t\bar{b}$ final state. 
This mode is characterized by two observable $b$-jets in the final state, and the $\sigma\cdot 
BR(W\to\ell\nu)$ is $1.1$ pb for each lepton flavor.

\section{Searches for single top quark at LHC}
\subsection{Overview of ATLAS searches}
In ATLAS, the three single-top channels have been studied in separate analyses. In all cases the $W$ boson from
the top-quark decay is required to decay leptonically, i.e.~in an $e$, $\mu$ or $\tau$, which, in the latter
case decays into an $e$ or $\mu$. 
The two analyses presented are carried out for early data and for an accumulated luminosity of 30 $\mathrm{fb}^{-1}$.

A common pre-selection has been developed for the three channels in order to discriminate against the dominant
backgrounds, $t\bar{t}$, $W+$jets and QCD multijet events. Selected events must have at least one isolated lepton,
i.e.~less than 6 GeV energy in a cone
with $\Delta R = 0.2$ around the lepton direction, with $p_T > 30$ GeV/$c$ in the central detector region.
The event is rejected if a second lepton is reconstructed with $p_T > 10$ GeV/$c$. Further, the presence of
at least two jets with $p_T > 30$ GeV/$c$ is required, while events are rejected if a fifth jet, when ordered in 
energy, is reconstructed with $p_T > 15$ GeV/$c$. Among the jets, the presence of a b-tagged jet with 
$p_T > 30$ GeV/$c$ is required. Finally, due to the leptonically decaying $W$, the transverse missing energy is required 
to be larger than 20 GeV.

\subsection{Overview of CMS searches}
All single top production modes have been studied in the CMS collaboration, considering leptonic final states for $t-$ and $s$-channel, 
and both the semi-leptonic and the di-leptonic decays in the $tW$ associated production~\cite{CMS.PTDR2}.

The CMS results summarized here are aimed to an integrated luminosity of $10-30$ fb$^{-1}$ with the nominal ``low luminosity'' LHC conditions ($\mathcal{L} = 2\cdot10^{33}\ \textrm{cm}^{-2}\textrm{s}^{-1}$, 25 ns spacing). No artificial misalignment has been introduced in the simulation, but the systematic uncertainties from the knowledge of the jet energy scale and b-tagging efficiency with 10 fb$^{-1}$ have been taken into account.

All analyses are based on simple counting of the events which pass some selection criteria, often normalizing the dominant backgrounds by counting events in a nearby region of the phase space.

\subsection{Event generators and detector simulation}
In ATLAS, Monte-Carlo simulation for all single-top signal was done using the AcerMC
matrix element generator interfaced to PYTHIA. For background processes the following are used: MC@NLO with HERWIG for
$t\bar{t}$; Alpgen with PYTHIA using K-factors from a comparison to a MCFM calculation for $W+$jets production,
properly including $Wc$, $Wcc$, $Wb$ and $Wbb$; and HERWIG for diboson production.\\

In CMS, signal samples have been generated using TopReX and SingleTop (based on CompHEP), with PYTHIA for the 
showering and hadronization steps; TopReX, Alpgen and PYTHIA have been used for the backgrounds.\\

Both experiments use full GEANT4 detector simulations for samples from the official Monte Carlo production, while fast
but accurate simulations (ATLFAST, FAMOS) were used for events not otherwise available, and in order to generate samples with
different parameters to investigate some of the systematic uncertainties.

\section{$t$-channel single top}
The $t$-channel final state for leptonic $W$ decays is characterized by one isolated lepton, 
missing transverse energy from the neutrino, one $b$-jet from the $t$ quark decay and 
one further jet with a broad $\eta$ spectrum extending up to $|\eta| \sim 4$.

\subsection{ATLAS}
In ATLAS two analysis strategies are pursued: a simple and robust cut-based analysis and a multivariate analysis
employing boosted decision trees (BDT).

\paragraph{Cut based analysis} In order to reduce the still abundant $W$+jets background the $b$-tagged jet is 
required to have a transverse momentum $p_T > 50$ GeV/$c$. The recoiling forward quark in the t-channel
produces a high $p_T$ light jet in the forward direction, therefore a
requirement on the hardest light jet of $|\eta| > 2.5$ 
reduces the $t\bar{t}$ background. The signal selection efficiency is $1.8\%$ and the S/B ratio is $0.37$ for an
integrated luminosity of 1 $\mathrm{fb}^{-1}$. It should be noted that the present result appears much less performant
when compared to earlier ATLAS publications~\cite{ATLASTDR}. The two reason are that Monte-Carlo generators have
evolved and in the TDR analysis some background processes have been neglected.
In particular: (i) the new PYTHIA
parton shower algorithm produces more energetic jets, (ii) for signal generation AcerMC has been used instead of
PYTHIA and for $W$+jets background Alpgen was used instead of HERWIG; (iii) the $t\bar{t}$ dileptonic channel, as
well as all channels involving $\tau$ decays were previously neglected.

\paragraph{Systematic uncertainties}
For a more detailed description of the nature of systematic uncertainties and the methods used to determine them
with Monte-Carlo simulation see~\cite{Gia}. Several systematic effects were considered and the 
largest contributions were found to arise from uncertainties in 
(i) the background cross sections, (ii) the jet energy scale, (iii) initial and final state radiation, 
(iv) the parton distribution functions, (v) limited Monte-Carlo statistics, (vi) b-tagging, (vii) the 
Monte-Carlo model.
Overall a total systematic uncertainty, including luminosity of $45\%$ ($22\%$) is estimated for  1 $\textrm{fb}^{-1}$
(10 $\textrm{fb}^{-1}$).

\paragraph{Multivariate analysis}
None of the above variables have a large discriminating power. Thus, a multivariate analysis in the form of
a boosted decision tree was trained against the $t\bar{t}$ background. About 40 variables were considered and
those which are least sensitive to the jet energy scale uncertainty were selected. For a full description of
the variables see~\cite{CSC}.

\paragraph{Uncertainty on the cross section measurement}
The estimated uncertainty on the cross section measurement using the BDT method above is
\begin{eqnarray*}
\frac{\Delta\sigma}{\sigma}(1\ \mathrm{fb}^{-1}) &\ =\ 5.7\%^{(stat)}\ \oplus\
20.2\%^{(syst)}\ \oplus\ 8.8\%^{(lumi)} &\simeq 23\%  \\
\frac{\Delta\sigma}{\sigma}(10\ \mathrm{fb}^{-1}) &\ =\ 1.8\%^{(stat)}\ \oplus\
8.5\%^{(syst)}\ \oplus\ 5.2\%^{(lumi)} &\simeq 10\%  \\
\end{eqnarray*}

\subsection{CMS}
The CMS analysis uses only the $W\to\mu\nu$ decays, which are expected to be cleaner. 
\paragraph{Event selection} 
Events are selected requiring a single isolated muon ($p_T > 19\ \textrm{GeV}/c$), $E_T^{miss} > 40\ \textrm{GeV}$, one central $b$-tagged jet ($P_{T} > 35\ \textrm{GeV}/c$) and one forward jet ($p_T > 40\ \textrm{GeV}/c$, $|\eta| > 2.5$). 
The purity of the selected sample was increased cutting on the transverse mass of the reconstructed $W$ boson, the mass of the top quark and the sum of the transverse momentum vectors of all reconstructed objects $\vec{\Sigma}_{T} = \vec{P}_{T,\ell} + \vec{E_{T}^{miss}} + \sum \vec{E}_{T,jet}$.
A full description of the reconstruction and selection can be found in \cite{CMS.NOTE.2006.084}.

\paragraph{Expected yield} The expected yield for an integrated luminosity of 10 fb$^{-1}$ is 2400 signal events and 1800 background events ($S/B \simeq 1.4$); the dominant backgrounds are $t\bar{t}$ ($67\%$) and $Wjj$ ($23\%$).


\paragraph{Uncertainty on the cross section measurement} The systematic uncertainty on the cross section has been computed taking into account theoretical uncertainties (PDFs, scale, $t$ and $b$ masses), the uncertainty on the overall jet energy scale ($5\%$ for $E_T(j) \le 25\ \textrm{GeV}$, $2.5\%$ for $E_T(j) > 50\ \textrm{GeV}$ and linearly interpolated in between), on the b-tagging efficiency ($4\%$) and on the total integrated luminosity ($5\%$).
The resulting uncertainty expected for $10\ \textrm{fb}^{-1}$ is
\[ \frac{\Delta\sigma}{\sigma}\ =\ 2.7\%^{(stat)}\ \oplus\ 8.0\%^{(syst)}\ \oplus\ 8.7\%^{(lumi)} \simeq 12\% \]
The contribution from the integrated luminosity is larger than $5\%$ because the subtraction of backgrounds not normalized using data increases the uncertainty as $\Delta\sigma/\sigma \simeq (1 + B/S) \cdot \Delta L/L$.\\

It is possible to perform a quick estimate of the expected uncertainty for an integrated luminosity of 1 fb$^{-1}$ by increasing the statistical uncertainty and scaling the systematic ones taking into account the expected knowledge of luminosity, jet energy scale and $b$-tagging efficiency for that luminosity (assuming they propagate linearly to $\Delta\sigma$); the signal is still expected to be observable with $\ge 5\sigma$ significance.

\section{$tW$ associated production}
The $tW$ associated production is a more challenging channel because a robust jet counting is necessary to reduce the large background 
from $t\bar{t}$.  Background normalization using control samples extracted from data is also necessary to avoid introducing large 
systematic uncertainties, as the achievable S/B ratio is rather poor.

\subsection{ATLAS}
Studies have focused on semi-leptonic final states ($Wt \to b\ell\nu qq$). Similarly to the t-channel analysis a baseline 
analysis using cuts and a multivariate analysis using BDTs have been performed.

\paragraph{Sequential cut analysis}
After pre-selection further requirements on b-jets are imposed. For the one b-tagged jet it is required that
$p_T > 50$ GeV/$c$. 
Further, a b-jet veto is applied: no further b-tagged jet above 35 GeV/$c$. Here, the b-tagging weight cut has been set by optimizing the signal
to background ratio and results in an efficiency of 30\%, while the standard b-tagging weight cut choice corresponds to an efficiency of 60\%.
Finally, if more than three high $p_T$ jets are present in the event, a $W$-mass constrain is imposed on the two leading untagged jets.
The resulting event yield for an integrated luminosity of $1 \mathrm{fb}^{-1}$ for final states with 2 (3, 4) jets 
(one of which is b-tagged) is $6360 \pm 230$ ($1090 \pm 80$, $380 \pm 40$) with a S/B ratio of 7\% (15\%, 11\%).

\paragraph{Multivariate analysis}
In order to discriminate against the different backgrounds, four BDT have been defined and optimized against the $t\bar{t}$ $\ell+$jets,
$t\bar{t}$ dileptonic, $W+$jets and the single top t-channel. As for the sequential cut analysis, the final states are separated according 
to the number of jets, while events with electrons and muons are combined. Thus, a total of 12 BDT were defined. 
A set of 25 discriminating variables
were identified, including opening angles between jets, pseudorapidity values, invariant (transverse) masses, jet $p_T$ and event shape
variables (see~\cite{CSC} for details).
With an optimal cut on the BDT outputs, the resulting event yields for the three final states 
are $170 \pm 10$ ($45 \pm 6$, $15.6 \pm 3.4$) with a S/B ratio of 35\% (46\%, 36\%).

\paragraph{Systematic uncertainties}
The $Wt$ analysis is affected by the imperfect knowledge of the background normalization, which will be determined using data when possible.
Also the modeling of the gluon radiation in the Monte-Carlo simulation adds a systematic uncertainty. On the detector side, significant
uncertainties are introduced through the use of the b-tagging algorithms (to select the first b-jet and to veto a second one). The b-tagging
and the jet energy scale uncertainties increase with the number of jets in the final state. We expect to obtain a $3 \sigma$ evidence
with a few $\mathrm{fb}^{-1}$ of data taking and estimate an uncertainty of 
\begin{eqnarray*}
\frac{\Delta\sigma}{\sigma}(1\ell) &\ =\ 6.6\%^{(stat)}\ \oplus\
17.7\%^{(syst)}\ \oplus\ 7.9\%^{(lumi)} &\simeq 20\% \\
\end{eqnarray*}
with 10 $\mathrm{fb}^{-1}$.

\subsection{CMS}
Two analysis strategies have been developed, one for the semi-leptonic events ($tW\to bWW \to b\ell\nu jj$) and one for the di-leptonic events ($tW \to bWW \to b\ell\nu\ell'\nu'$)\cite{CMS.NOTE.2006.086}.

Information from the reconstructed tracks and jet shape variables were used in order to make jet counting more robust against calorimetric noise and pile-up.

\paragraph{Semi-leptonic selection} In the semi-leptonic decays, events are selected requiring one isolated lepton and exactly three jets, exactly one of which must be $b$-tagged; a cut on the missing transverse energy has also been applied to suppress $QCD$ background. 

Additional selection cuts are applied to the transverse mass of the leptonically decaying $W$, on the mass of the two light jets and on the reconstructed top quark mass; the $b\!\leftrightarrow\!W$ pairing was done combining in a Fisher discriminator angular variables, $p_T(top)$ and the charges (using for the jets the $p_T$ weighted sum of the charges of the tracks in the jet cone).

A control sample dominated by $t\bar{t}$ events was selected by requiring one additional jet.

\paragraph{Di-leptonic selection} Two high $p_T$ isolated leptons are required, opposite in flavor ($e+\mu$) to suppress the 
$Z/\gamma^* \to \ell^+\ell^-$ process, and a single jet, $b$-tagged. 

To normalize the dominant background from di-leptonic $t\bar{t}$ a control sample was selected requiring one additional jet; the jet is required to be $b$-tagged to avoid signal contamination in this sample.

\paragraph{Expected yields} The expectations in the semi-leptonic channel are 1700 signal events and 8700 background events for 
10 fb$^{-1}$, with S/B $\simeq 0.2$; the dominant background is $t\bar{t}$ ($83\%$). 
In the di-leptonic case the expected yield is 560 signal events and 1500 background events ($96\%$ $t\bar{t}$), corresponding to S/B $\simeq 0.4$.

\paragraph{Uncertainty on the cross section measurements} Many sources of systematic uncertainties have been considered, both theoretical (PDFs, background cross sections) and experimental (jet energy scale, b-tagging efficiency, amount of pile-up). 
The expected accuracy on the cross sections is:
\begin{eqnarray*}
\frac{\Delta\sigma}{\sigma}(1\ell) &\ =\ 7.5\%^{(stat)}\ \oplus\ 15.6\%^{(syst)}\ \oplus\ 7.8\%^{(lumi)} &\simeq 19\% \\
\frac{\Delta\sigma}{\sigma}(2\ell) &\ =\ 8.8\%^{(stat)}\ \oplus\ 22.8\%^{(syst)}\ \oplus\ 5.4\%^{(lumi)} &\simeq 25\% 
\end{eqnarray*}
The dominant systematic uncertainties are from the jet energy scale (both channels), b-tagging efficiency (di-leptonic) and amount of pile-up (semi-leptonic).

\section{$s$-channel single top}
At LHC the expected $s$-channel single top cross section is 10 pb, an order of magnitude smaller than $t$-channel and almost a factor 100 smaller than $t\bar{t}$, so top backgrounds are much more an issue than at the Tevatron where $\sigma_{s-ch} \!\sim\! 1$ pb,  $\sigma_{t-ch} \!\sim\! 2$ pb,  $\sigma_{t\bar{t}} \!\sim\! 7$ pb.

\subsection{ATLAS}
A default cut-based selection and a likelihood-based analysis have been done and are outlined in the following.

\paragraph{Sequential cut analysis}
After event pre-selection cuts are defined to take advantage of the specific characteristics of the s-channel. Exactly two $b$-tagged jets are
required and no further jet above 15 GeV/$c$ is allowed. This reduces the $t\bar{t}$, $W+$jets and QCD multijet contamination. The selection
requires also: the two-jet opening angle to be $0.5 < \Delta R(b,b) < 4.0$, the scalar sum of the total jet transverse momenta $H_T(\mathrm{jet})$
to be $80 < H_T(\mathrm{jet}) < 220$ GeV and the sum of the transverse missing energy and lepton transverse momentum to be ranging 
between $60 < E_T^{miss} + p_T(\ell) < 130$ GeV.

With this selection the overall signal efficiency is $1.1\%$, yielding 25 events for 1 $\mathrm{fb}^{-1}$ with a S/B ratio of 10\%.

\paragraph{Likelihood selection}
The above analysis shows that the s-channel is plagued by a high level of background. Therefore an analysis based on the likelihood 
method has been developed, starting from the pre-selection and the two b-jet requirements. Out of a large pool of variables the
16 most discriminating variables were chosen, including opening angles, pseudorapidity values, invariant masses, transverse masses of the
reconstructed top candidates, $H_T$ and event shape variables. It is assumed that the shape of the distributions are well known and
validated on data itself. Five likelihood functions are defined, each designed to discriminate against a given background: 
$t\bar{t}\;\ell+$jets, $t\bar{t}$ dileptonic, $t\bar{t}\;\tau+\ell$, $W+$jets, t-channel. On each of the likelihood functions a 
simple threshold is applied in order to minimize the uncertainty in the cross-section measurement.
The likelihood method exhibits a similar statistical sensitivity to the cut analysis with an improved signal-to-background ratio of 20\%.

\paragraph{Systematic uncertainties}
The relatively poor S/B ratio makes measurements in this channel a real challenge. Even more important than in the other
channels is the background normalization, the b-tagging efficiency and the jet energy scale. The effect of ISR/FSR will 
need to be studied from data samples. Overall we estimate an uncertainty in the cross section for 10 $\mathrm{fb}^{-1}$ of

\[ \frac{\Delta\sigma}{\sigma} \ =\ 20\%^{(stat)}\ \oplus\
45\%^{(syst)}\ \oplus\ 18\%^{(lumi)} \simeq 52\% \]

\subsection{CMS}
An analysis has been done using $W\to e\nu$ and $W\to\mu\nu$ events together, and requiring two $b$ tags.\cite{CMS.NOTE.2006.084}

\paragraph{Event reconstruction and selection}
The $s$-channel selection requires one isolated high $p_T$ lepton ($p_T > 19\ \textrm{GeV}/c$, $|\eta(e)| < 2.4$, $|\eta(\mu)| < 2.1$), two jets ($|\eta| < 2.5$, $E_{T} > 50\ \textrm{GeV}$) both tagged as $b$, $E_T^{miss} > 30 \ \textrm{GeV}$; there must be no other jets (with $E_T > 20\ \textrm{GeV}$) nor leptons (with $p_T > 10\ \textrm{GeV}/c$). 

The $W$ boson is reconstructed from the lepton and the missing transverse energy by imposing the mass constraint, and the top quark is reconstructed pairing the $W$ with the $b$-jet with most opposite ``jet charge'' (the $p_T$ weighted sum of the charges of the tracks in the jet cone).

After the reconstruction, additional cuts are applied to the kinematic observables $M_T(W)$, $M(t)$, $|\vec{\Sigma}_T|$ and the total hardness $H_{T}$ (scalar sum of the $E_T$ of all reconstructed objects including the neutrino).

\paragraph{Expected yields}
For an integrated luminosity of 10 fb$^{-1}$ the expected number of signal events passing all selection cuts is 270, 
together with 2000 background events (dominant contributions: $62\%$ $t\bar{t}$, $31\%$ $t$-channel single top), for a S/B $\simeq 0.13$.

\paragraph{Systematic uncertainties} In order to reduce the systematic uncertainties on the background normalization, two control samples extracted from data are used, one enriched in semi-leptonic $t\bar{t}$ and one in di-leptonic $t\bar{t}$. Nevertheless, there is still a large systematic uncertainty, mostly arising from the knowledge of the jet energy scale.

\[ \frac{\Delta\sigma}{\sigma} \ =\ 18\%^{(stat)}\ \oplus\ 31\%^{(syst)}\ \oplus\ 19\%^{(lumi)} \simeq 41\% \]

\section{Studies on QCD multi-jet background}
The cross section for multi-jet QCD production is many order of magnitude larger than the one for signal, but the selection efficiency is expected to be very small; as it's not practically possible to simulate a number of events corresponding to integrated luminosities of some fb$^{-1}$, the efficiency cannot be determined just by counting how many simulated events pass the selection. 

\paragraph{ATLAS}
By requiring the presence of one isolated lepton and one $b$-jet the QCD background is reduced very effectively. However, the uncertainty
on the remaining background might be rather large. Thus, the QCD background contamination in selected samples will be monitored, e.g.~by
reconstructing the $W$ transverse mass. By using the distribution in the region below 50 GeV one can normalize the level of
remaining background and assess the expected level in the signal region above 50 GeV. For the studies presented here no 
uncertainty is attributed to this background.

\paragraph{CMS}
In order to provide an estimate of the yield of this background the ``combined efficiency'' method was used in CMS analyses. 

The first steps of the selection have been factorized in sets of approximatively independent cuts; for each set the efficiency on multi-jet events was determined directly on simulated events; the combined efficiency of all cut sets was estimated as product of the efficiencies of all sets.

Eventually, the total selection efficiency was obtained starting from the above estimate and using the efficiency for signal events as a conservative limit on the efficiency for the background for the final selection cuts.

The pairwise correlation of individual cut sets has also been checked with simulated data, and within the available statistics the approximation was found to be good.

According to these estimates, multi-jet background is negligible in all analysis channels except semi-leptonic $Wt$, where $B_{QCD}/S \sim 30\%$.

\section{Conclusions}
All single top production channels have been studied by the two experiments, and the expected uncertainties are in reasonable agreement.
The $t$-channel production, thanks to the large expected cross section (240 pb) should be observable with the first few 
fb$^{-1}$ of integrated luminosity.
The $Wt$ associated production, invisible at Tevatron ($\sigma \simeq
0.2$ pb) but sizeable at LHC ($\sigma \simeq 66$ pb), 
is more challenging but should be in reach with 10-20 fb$^{-1}$. 
Observing the $s$-channel production appears to be much harder, and will probably require even more data and advanced analysis techniques.

\end{document}